\documentclass[prc,superscriptaddress,noshowpacs,unsortedaddress,twocolumn,showpacs,preprintnumbers,amsmath,amssymb]{revtex4-1}

\usepackage[dvipdfmx]{graphicx}
\usepackage{amsmath,amssymb,times}
\usepackage{color}
\usepackage{ulem}
\usepackage{bm}
\usepackage{here}

\def\lsim{~\,\makebox(1,1){$\stackrel{<}{\widetilde{}}$}\,~}

\newcommand{\beq}{\begin{equation}}
\newcommand{\eeq}{\end{equation}}
\newcommand{\bea}{\begin{eqnarray}}
\newcommand{\eea}{\end{eqnarray}}

\newcommand{\bfi}[1]{\mbox{\boldmath $#1$}}

\newcommand{\vK}{{\bfi K}}

\newcommand{\vs}{{\bfi s}}

\newcommand{\vrr}{{\bfi r}}
\newcommand{\vR}{{\bfi R}}

\begin{document}

%\preprint{}

\title{Neutron skin thickness of ${}^{208}$Pb determined from reaction cross section 
for proton scattering
}



\author{Shingo~Tagami}
%\email[]{sh.tagami@gmail.com}
\author{Tomotsugu~Wakasa}
%\email[]{wakasa@phys.kyushu-u.ac.jp}    
\author{Jun~Matsui}
%\email[]{j.matsui@cmt.phys.kyushu-u.ac.jp}
\author{Masanobu Yahiro}
\email[]{orion093g@gmail.com}
\affiliation{Department of Physics, Kyushu University, Fukuoka 819-0395, Japan} 
\author{Maya~Takechi}
%\email[]{takechi@np.gs.niigata-u.ac.jp}    
\affiliation{Niigata University, Niigata 950-2181, Japan}

%\thanks{}
%%%\affiliation{}


\date{\today}

\begin{abstract}
\begin{description}
\item[Background]
 The reaction cross section $\sigma_R$ is useful 
to determine the neutron radius $R_n$ as well as the 
matter radius $R_m$.
 The chiral (Kyushu) $g$-matrix folding model for $^{12}$C scattering 
on  $^{9}$Be, $^{12}$C, $^{27}$Al targets was tested in 
the incident energy range of $30  \lsim E_{\rm in} \lsim 400 $ MeV, 
and it is found that the model reliably reproduces the $\sigma_R$ 
in $30  \lsim E_{\rm in} \lsim 100 $ MeV and 
$250  \lsim E_{\rm in} \lsim 400$ MeV.  
\item[Aim]
 We determine $R_n$ and the neutron skin thickness $R_{\rm skin}$ of 
${}^{208}{\rm Pb}$ by using high-quality $\sigma_R$ data for the 
$p+{}^{208}{\rm Pb}$ scattering in $30  \leq E_{\rm in} \leq 100$ MeV.
 The theoretical model is the Kyushu $g$-matrix folding model  with the  densities 
calculated with Gongny-D1S HFB (GHFB) with the angular momentum  projection (AMP). 
\item[Results]
 The Kyushu $g$-matrix folding model with the GHFB+AMP densities 
underestimates $\sigma_{\rm R}$ in $30 \leq E_{\rm in} \leq 100$~MeV only by a factor of 0.97.  
 Since the proton radius $R_p$ calculated with GHFB+AMP agrees with the precise experimental data of 5.444 fm, 
 the small deviation of the theoretical result from the data on $\sigma_R$  
 allows us to scale the  GHFB+AMP neutron density so as to reproduce the $\sigma_R$ data. 
 In $E_{\rm in}$ = 30--100 MeV, the experimental $\sigma_R$ data can be 
reproduced by assuming the neutron radius of ${}^{208}{\rm Pb}$ as 
$R_n$ = $5.722 \pm 0.035$ fm.
\item[Conclusion]
 The present result $R_{\rm skin}$ = $0.278 \pm 0.035$ fm is in good agreement with the recent PREX-II result 
of $r_{\rm skin}$ = $0.283\pm 0.071$ fm.
\end{description}
\end{abstract}

%\keywords{}

\maketitle

\clearpage

\section{Introduction}
\label{Introduction}

 Horowitz {\it et al.} \cite{PRC.63.025501} proposed a direct measurement 
for neutron skin $R_{\rm skin}$ = $R_n - R_p$, where 
$R_n\equiv\langle r_n^2\rangle^{1/2}$ and 
$R_p\equiv\langle r_p^2\rangle^{1/2}$ 
are the root-mean-square (rms) radii of point neutrons  
and protons, respectively.
 The measurement consists of parity-violating ($PV$) and 
elastic electron scattering. 
 The neutron radius $R_n$ is determined from the former experiment, whereas
the proton radius $R_p$ is from the latter.

 Very recently, by combining the original Lead Radius EXperiment (PREX) 
result \cite{PRL.108.112502,PRC.85.032501} with the updated PREX-II result, 
the PREX collaboration reported the following value~\cite{PREX-II2021}:
\begin{equation}
R_{\rm skin}^{PV} = 0.283\pm 0.071\,{\rm fm},
\end{equation}
where the quoted uncertainty represents a $1\sigma$ error 
and has been greatly reduced from the original value of
$\pm 0.177$ fm (quadrutic sum of experimental and model uncertainties) 
\cite{PRC.85.032501}.
 The $R_{\rm skin}^{PV}$ value is most reliable at the present 
stage, and provides crucial tests for the equation of state (EoS) 
of nuclear matter 
\cite{PRC.102.051303,AJ.891.148,AP.411.167992,%
EPJA.56.63,JPG.46.093003}
as well as nuclear structure models.
 For example, Reed {\it et al.} \cite{arXiv.2101.03193} 
report a value of the sloop parameter $L$ of the EoS 
and examine the impact of such a stiff symmetry energy 
on some critical neutron-star observables.
 It should be noted that the $R_{\rm skin}^{PV}$ value 
is considerably larger than the other experimental 
values which are significantly model dependent 
\cite{PRL.87.082501,PRC.82.044611,PRL.107.062502,%
PRL.112.242502}.
 As an exceptional case, a nonlocal dispersive-optical-model 
(DOM) analysis of ${}^{208}{\rm Pb}$ deduces 
$r_{\rm skin}^{\rm DOM} =0.25 \pm 0.05$ fm \cite{PRC.101.044303}, 
which is consistent with $R_{\rm skin}^{PV}$.
 It is the aim of this paper to present the 
$R_{\rm skin}$ value with a similar precision of $R_{\rm skin}^{PV}$ 
by analyzing the reaction cross section $\sigma_R$ 
for $p+{}^{208}{\rm Pb}$.

 The reaction cross section $\sigma_R$ is a powerful tool 
to determine matter radius  $R_m$. 
 One can evaluate $R_{\rm skin}$ and $R_n$ by using 
the $R_m$ and the $R_p$ \cite{ADNDT.99.69} determined by the electron scattering. 
 The $g$-matrix folding model is a standard way of 
deriving microscopic optical potential 
for not only proton scattering but also 
nucleus-nucleus scattering
\cite{NPA.291.299,*NPA.291.317,*NPA.297.206,PR.55.183,*Satchler83,PTP.70.459,*PTP.73.512,*PTP.76.1289,ANP.25.275,%
PRC.78.044610,*PRC.79.011601,*PRC.80.044614,PRC.89.064611,JPG.42.025104,*JPG.44.079502,PRC.92.024618,*PRC.96.059905,%
PTEP.2018.023D03,PRC.101.014620,PRL.108.052503}.
 Applying the folding model with the Melbourne $g$-matrix~\cite{ANP.25.275} 
for  interaction cross sections $\sigma_{\rm I}$ for Ne isotopes and $\sigma_{\rm R}$ for Mg isotopes, 
we discovered that $^{31}$Ne is a halo nucleus with large deformation~\cite{PRL.108.052503},   
and deduced the matter radii $r_{\rm m}$ for Ne isotopes~\cite{PRC.85.064613} and  
for Mg isotopes~\cite{PRC.89.044610}. 
 The folding potential is nonlocal, but is localized with the method of 
Ref.~\cite{NPA.291.299,*NPA.291.317,*NPA.297.206}. 
 The validity is shown in Ref.~\cite{JPG.37.085011}.
 For proton scattering, the localized version of $g$-matrix folding
model \cite{PRC.88.054602} yields the same results 
as the full folding $g$-matrix folding model of Ref.~\cite{ANP.25.275}, 
as shown by comparing the results of Ref.~\cite{PRC.88.054602} 
with those of Ref.~\cite{ANP.25.275}.

\begin{figure}[tbp]
\begin{center}
 \includegraphics[width=0.45\textwidth,clip]{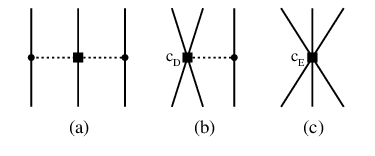}
 \caption{3$N$Fs in NNLO. 
Diagram (a) corresponds 
to the Fujita-Miyazawa 2$\pi$-exchange 3$N$F \cite{PTP.17.360,*PTP.17.366}, 
and diagrams (b) and (c) correspond to 1$\pi$-exchange and contact 3$N$Fs.
 The solid and dashed lines denote nucleon and pion propagations, 
respectively, and filled circles and squares stand for vertices. 
 The strength of the filled-square vertex is often called $c_{D}$ 
in diagram (b) and $c_{E}$ in diagram (c). 
}
 \label{fig:diagram}
\end{center}
\end{figure}

 Recently, Kohno \cite{PRC.88.064005,*PRC.96.059903} calculated the $g$-matrix 
for the symmetric nuclear matter, 
using the Brueckner-Hartree-Fock method with chiral
4th-order (${\rm N^3LO}$) nucleon-nucleon ($NN$) forces (2$N$Fs) and
3rd-order (NNLO) three-nucleon forces (3$N$Fs).
 He set $c_D=-2.5$ and $c_E=0.25$ so that  the energy per nucleon can  become minimum 
at $\rho = \rho_{0}$; see Fig.~\ref{fig:diagram} for $c_{D}$ and $c_{E}$.
 Toyokawa {\it et al.} \cite{PTEP.2018.023D03} localized 
the non-local chiral  $g$-matrix into three-range Gaussian forms.
using the localization method proposed 
by the Melbourne group~\cite{ANP.25.275,PRC.44.73,PRC.49.1309}. 
The resulting local $g$-matrix is called  ``Kyushu  $g$-matrix''.

%----------------------
% Figure Ch-3NF diagram
%----------------------
%----------------------

 The  Kyushu $g$-matrix folding model is successful in reproducing $\sigma_{\rm R}$ 
and differential cross sections  $d\sigma/d\Omega$ for $^4$He scattering 
in $E_{\rm in}$ = 30--200 MeV/nucleon \cite{PTEP.2018.023D03}. 
 The success is true for proton scattering at $E_{\rm in}$ = 65 MeV 
\cite{JPG.42.025104,*JPG.44.079502}. 
 Lately, we predicted neutron skin $r_{\rm skin}$ and  proton, neutron,
matter radii, $R_p$, $R_n$, $R_m$ from 
interaction cross sections $\sigma_{\rm I}$ ($\approx \sigma_{\rm R}$)  
for $^{42-51}$Ca+$^{12}$C scattering at $E_{\rm in}$ = 280 MeV/nucleon,
using the Kyushu $g$-matrix folding model with 
the densities calculated with Gongny-D1S HFB (GHFB) 
with and without the angular momentum  projection (AMP) \cite{PRC.101.014620}.

 In Ref.~\cite{PRC.101.014620}, we tested the Kyushu $g$-matrix folding model for $^{12}$C scattering 
on  $^{9}$Be, $^{12}$C, $^{27}$Al targets in  $30  \lsim E_{\rm in} \lsim 400 $~MeV, comparing the theoretical 
$\sigma_{\rm R}$ with the experimental data \cite{PRC.79.061601}. 
 We found that the Kyushu $g$-matrix folding model is reliable for $\sigma_{\rm R}$ 
in $30  \lsim E_{\rm in} \lsim 100 $~MeV and $250  \lsim E_{\rm in} \lsim 400 $~MeV.  
 This indicates that the Kyushu $g$-matrix folding model is applicable in $30  \le E_{\rm lab} \le 100$~MeV, although 
the data on p+$^{208}$Pb scattering are available in $21  \le E_{\rm lab} \le 180$~MeV.
 
 In this paper, we present the determination of 
$R_{\rm skin}^{\rm GHFB}$ from the measured $\sigma_R$ 
for $p+{}^{208}{\rm Pb}$ scattering in 
$30  \leq E_{\rm in} \leq 100$ MeV 
\cite{PRC.12.1167,NPA.653.341,PRC.71.064606}, 
using the Kyushu $g$-matrix folding model  with the GHFB+AMP densities. 
 As mentioned above, the Kyushu $g$-matrix folding model is applicable in $30  \le E_{\rm in} \le 100$~MeV, 
although the data on $p+{}^{208}{\rm Pb}$ scattering are available
in $21  \le E_{\rm in} \le 180$~MeV.  
 In Sec.~\ref{sec:model}, we briefly describe our model.
 Section~\ref{sec:results} presents the results and a comparison 
with $R_{\rm skin}^{PV}$, and discussion follows.
Finally, Sec.~\ref{sec:summary} is devoted to a summary.

\section{Model}
\label{sec:model}

 Our model is the Kyushu $g$-matrix folding model~\cite{PTEP.2018.023D03} 
with the densities calculated with GHFB+AMP \cite{PRC.101.014620}.  
 In Ref.~\cite{PTEP.2018.023D03}, the Kyushu $g$-matrix is constructed 
from chiral interaction with the cutoff $\Lambda$ = 550 MeV.  
 The model was tested for $^{12}$C scattering 
on $^{9}$Be, $^{12}$C, and $^{27}$Al targets in  $30  \lsim E_{\rm in} \lsim 400 $~MeV. 
 It is found that the Kyushu $g$-matrix folding model is good 
in $30 \lsim E_{\rm in} \lsim 100 $~MeV and 
$250  \lsim E_{\rm in} \lsim 400$~MeV \cite{PRC.101.014620}.  

 The brief formulation of the folding model itself is shown below. 
 For nucleon-nucleus scattering, the potential is composed of the direct and exchange parts,
$U^{\rm DR}$ and $U^{\rm EX}$~\cite{PRC.89.044610}:
\begin{subequations}
\begin{eqnarray}
U^{\rm DR}(\vR) & = & 
\sum_{\mu,\nu}\int             \rho^{\nu}_{\rm T}(\vrr_{\rm T})
            g^{\rm DR}_{\mu\nu}(s;\rho_{\mu\nu})  d
	    \vrr_{\rm T}\ ,\label{eq:UD} \\
U^{\rm EX}(\vR) & = & 
\sum_{\mu,\nu}
\int \rho^{\nu}_{\rm T}(\vrr_{\rm T},\vrr_{\rm T}+\vs) \nonumber \\
                &   &
\times g^{\rm EX}_{\mu\nu}(s;\rho_{\mu\nu}) \exp{[-i\vK(\vR) \cdot \vs/M]}
             d \vrr_{\rm T}\ ,\label{eq:UEX}
\end{eqnarray}
\end{subequations}
where $\vR$ is the relative coordinate between a projectile (P)  and 
a target (${\rm T}$),
$\vs=-\vrr_{\rm T}+\vR$, and $\vrr_{\rm T}$ is
the coordinate of the interacting nucleon from T.
 Each of $\mu$ and $\nu$ denotes the $z$-component of isospin; 
$1/2$ means neutron and $-1/2$ does proton.
 The nonlocal $U^{\rm EX}$ has been localized in Eq.~\eqref{eq:UEX}
with the local semi-classical approximation
\cite{NPA.291.299,*NPA.291.317,*NPA.297.206},
where \vK(\vR) is the local momentum between P and T, 
and $M= A/(1 +A)$ for the target mass number $A$;
see Ref.~\cite{JPG.37.085011} for the validity of the localization.
 The direct and exchange parts, $g^{\rm DR}_{\mu\nu}$ and
$g^{\rm EX}_{\mu\nu}$, of the $g$-matrix depend on the local density
\bea
 \rho_{\mu\nu}=\rho^{\nu}_{\rm T}(\vrr_{\rm T}+\vs/2)\ ,
\label{local-density approximation}
\eea
at the midpoint of the interacting nucleon pair; see
Ref.~\cite{PRC.85.064613} for the explicit forms of $g^{\rm DR}_{\mu\nu}$ and
$g^{\rm EX}_{\mu\nu}$.

 The relative wave function $\psi$ is decomposed into partial waves $\chi_L$,
each with different orbital angular momentum $L$.
 The elastic $S$-matrix elements $S_L$ are obtained 
from the asymptotic form of the $\chi_L$.
 The total reaction cross section $\sigma_{\rm R}$ is calculable 
from the $S_L$ as
\bea
\sigma_{\rm R}=\frac{\pi}{K^2}\sum_L (2L+1)(1-|S_L|^2)\ .
\eea

 The proton and neutron densities, $\rho_p(r)$ and $\rho_n(r)$, 
are calculated with GHFB+AMP. 
 As a way of taking the center-of-mass correction to the densities, 
we use the method of Ref.~\cite{PRC.85.064613}, 
since the procedure is quite simple. 

%Results
\section{Results}
\label{sec:results} 

\begin{figure}
\begin{center}
  \includegraphics[width=0.45\textwidth,clip]{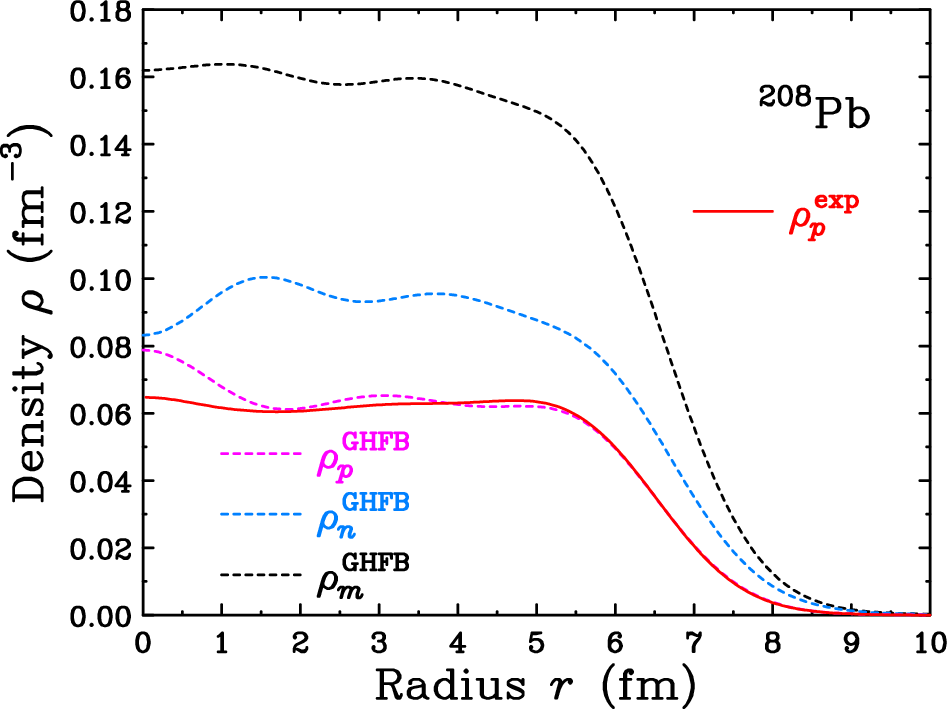}
 \caption{ 
 $r$ dependence of densities,  $\rho_p(r)$, $\rho_n(r)$, $\rho_m(r)$, for $^{208}$Pb calculated with 
 GHFB+AMP.  Three dashed  lines from the bottom to the top denote $\rho_p(r)$, $\rho_n(r)$, $\rho_m(r)$, respectively. 
 The experimental point-proton (unfolded) density $\rho_p$
is taken from Refs.~\cite{ADNDT.36.495,NPA.298.452}.
   }
 \label{Fig-Densities+Pb}
\end{center}
\end{figure}

 Figure~\ref{Fig-Densities+Pb} shows the 
proton $\rho_p^{\rm GHFB}$,
neutron $\rho_n^{\rm GHFB}$, and 
matter  $\rho_m^{\rm GHFB}\equiv \rho_p^{\rm GHFB}+\rho_n^{\rm GHFB}$
densities as a function of $r$.
 The experimental point-proton distribution extracted from 
the electron scattering data is also shown.
 The theoretical proton distribution $\rho_p^{\rm GHFB}$ 
reproduces the experimental $\rho_p^{\rm exp}$ reasonably well.

%%%%%%%%%%%%%%%%%%%%%%%
%%%  Figure
%%%%%%%%%%%%%%%%%%%%%%%

\begin{figure}
\begin{center}
 \includegraphics[width=0.45\textwidth,clip]{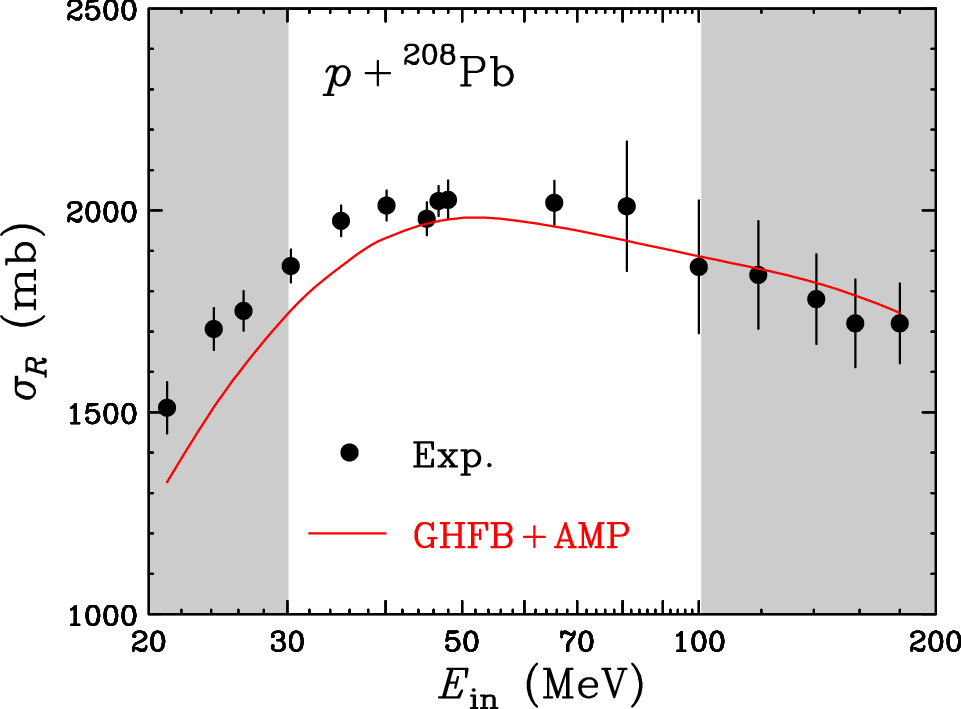}
 \caption{ 
 $E_{\rm in}$ dependence of reaction cross sections $\sigma_{\rm R}$ 
 for $p$+$^{208}$Pb scattering. 
 The solid line stands for the results of the Kyushu $g$-matrix  folding model with GHFB+AMP densities. 
 The data are taken from Refs.~\cite{PRC.12.1167,NPA.653.341,PRC.71.064606}. 
   }
 \label{Fig-RXsec-p+Pb-1}
\end{center}
\end{figure}

 The Kyushu $g$-matrix folding model with the GHFB+AMP densities 
underestimates the $\sigma_R$ data
in $30 \leq E_{\rm in} \leq 100$ MeV only 
by a factor of 0.97, as shown in Fig.~\ref{Fig-RXsec-p+Pb-1}.  
 The proton radius $R_p^{\rm GHFB}$ = 5.444 fm calculated with GHFB+AMP agrees with 
 the experimental value of $R_p^{\rm exp}$ = 5.444 fm \cite{PRC.90.067304}.
 Because of $\sigma_R \propto R_m^2$, 
the observed discrepancy of $\sigma_R$ is attributed 
to the underestimation of $\rho_m^{\rm GHFB}$ originating 
from the underestimation of $\rho_n^{\rm GHFB}$.
 Small deviation makes it possible to scale the GHFB+AMP densities for the neutron density 
so as to reproduce $\sigma_R^{\rm exp}$ in $E_{\rm in}$ = 30--100 MeV.
The result of the scaling is $R_n^{\rm exp}=5.722 \pm 0.035$~fm leading to 
\bea
R_{\rm skin}^{\rm exp}=0.278 \pm 0.035~{\rm fm}. 
\label{eq:final-skin}
\eea
This  result is consistent with $R_{\rm skin}^{PV}$ 
= $0.283\pm 0.071$~fm.
 
%%%%%%%%%%%%%%%%%%%%%%%
%%%  Figure
%%%%%%%%%%%%%%%%%%%%%%%

Now we show a simple derivation of $R_n^{\rm exp}$ in the limit of $K^{\rm exp}=K^{\rm th}$. 
The experimental and theoretical (GHFB+AMP) reaction cross sections,
$\sigma_R^{\rm exp}$ and $\sigma_R^{\rm th}$, can be expressed 
as 
\begin{subequations}
\begin{eqnarray}
\sigma_R^{\rm exp} & = & 
K^{\rm exp}
\left[
 (R_p^{\rm exp})^2\frac{Z}{A}
+(R_n^{\rm exp})^2\frac{N}{A}
\right]\ ,\\
\sigma_R^{\rm th} & = & 
K^{\rm th}
\left[
 (R_p^{\rm th})^2\frac{Z}{A}
+(R_n^{\rm th})^2\frac{N}{A}
\right]\ ,
\end{eqnarray}
\end{subequations}
where $Z$, $N$, and $A$ are proton, neutron, and atomic 
numbers of ${}^{208}{\rm Pb}$, respectively, and $K$ is 
a proportional coefficient between $\sigma_R$ and 
$R_m^2=R_p^2(Z/A)+R_n^2(N/A)$.
By using $K^{\rm exp}=K^{\rm th}$ and $R_p^{\rm exp}=R_p^{\rm th}$, the experimental neutron 
radius $R_n^{\rm exp}$ can be deduced as 
\begin{equation}
R_n^{\rm exp} 
=
\sqrt{
\frac{Z(R_p^{\rm exp})^2+N(R_n^{\rm th})^2}{N\sigma_R^{\rm th}}
\sigma_R^{\rm exp}
-
(\sigma_p^{\rm exp})^2\frac{Z}{N}
}\ ,
\label{Eq:Rn-limit}
\end{equation}
from the experimental $\sigma_R^{\rm exp}$ and $R_p^{\rm exp}$ data 
and the theoretical $R_n^{\rm th}$ in GHFB+AMP.

\begin{figure}
\begin{center}
 \includegraphics[width=0.45\textwidth,clip]{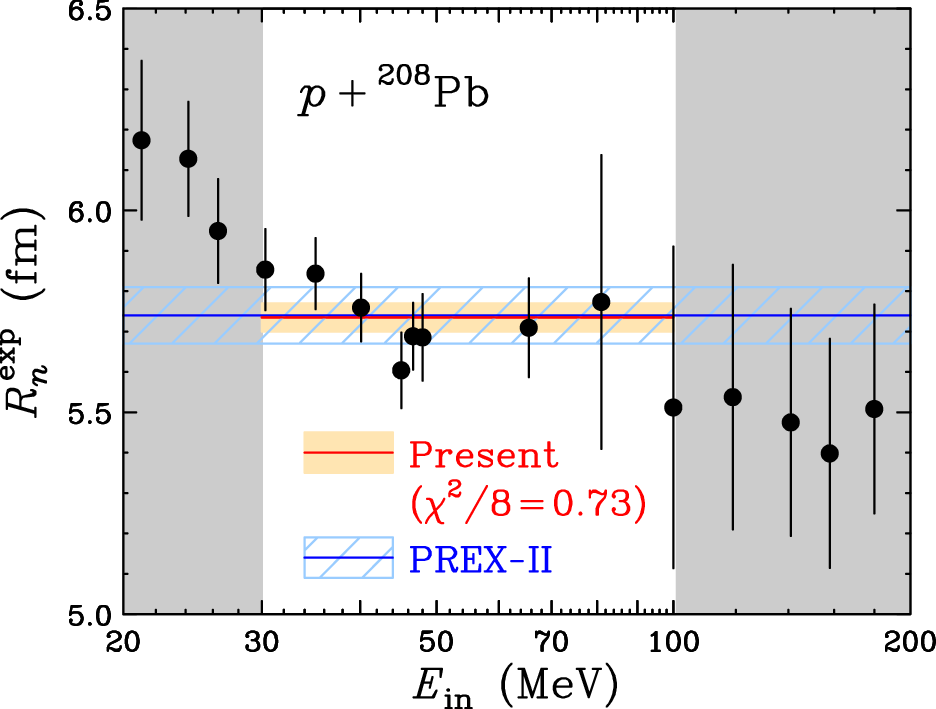}
 \caption{ 
 Neutron radius $R_n^{\rm exp}$ of ${}^{208}{\rm Pb}$ deduced 
from the $p+{}^{208}{\rm Pb}$ reaction cross section and 
the theoretical Kyushu $g$-matrix folding model calculations 
as a function of infident energy $E_{\rm in}$.
   }
 \label{Fig-Rn-Edep}
\end{center}
\end{figure}

 Figure~\ref{Fig-Rn-Edep} shows the $R_n^{\rm exp}$ results 
as a function of incident energy $E_{\rm in}$.
 The deduced $R_n^{\rm exp}$ values are almost 
 independent of $E_{\rm in}$ 
in the region of $E_{\rm in}$ = 30--100 MeV where 
the present folding model is reliable~\cite{PRC.101.014620}.
By combining the eight data in this energy region, 
the neutron radius of ${}^{208}{\rm Pb}$ becomes 
$\overline{R}_n^{\rm exp} = 5.735\pm 0.035$ fm
as shown by the filled band in Fig.~\ref{Fig-Rn-Edep}.
 This result shows that the neutron skin thickness of 
${}^{208}{\rm Pb}$ is $R_{\rm skin}^{\rm exp} = 0.291\pm 0.035$ fm 
with $R_p^{\rm exp}$ = 5.444 fm \cite{PRC.90.067304}. 
The limit of $K^{\rm exp}=K_R^{\rm th}$ is thus good, 
since $R_{\rm skin}^{\rm exp} = 0.291\pm 0.035$ fm is close to Eq.\eqref{eq:final-skin}. 
Equation \eqref{Eq:Rn-limit} is 
quite useful when $\sigma_R^{\rm exp} \approx \sigma_R^{\rm th}$ and 
$R_p^{\rm exp} \approx R_p^{\rm th}$.

%%%%%%%%%%%%%%%%%%%%%%%
%%%  Figure
%%%%%%%%%%%%%%%%%%%%%%%

%Summary
\section{Summary}
\label{sec:summary} 

 The proton radius $R_p$ calculated with GHFB+AMP agrees with the precise experimental data of 5.444 fm.
 In $30  \le E_{\rm in} \le 100$~MeV, 
we can obtain  $r_{\rm n}^{\rm exp}$ from $\sigma_{\rm R}^{\rm exp}$ 
by scaling the GHFB+AMP neutron density so as to reproduce $\sigma_{\rm R}^{\rm exp}$ for each $E_{\rm in}$,  
and take the weighted mean and its error for the resulting $r_{\rm n}^{\rm exp}$. 
 From the resulting $R_n^{\rm exp}=5.722 \pm 0.035$~fm and $r_{\rm p}^{\rm exp}=5.444$~{\rm fm},
we can get $R_{\rm skin}^{\rm exp}=0.278 \pm 0.035~{\rm fm}$. 

 In conclusion, our result $R_{\rm skin}^{\rm exp}=0.278 \pm 0.035$~fm is consistent with a new result 
$r_{\rm skin}^{208}({\rm PREX~II}) =0.283 \pm 0.071~{\rm fm}$ of PREX-II. 

\begin{acknowledgments}
We would like to thank Dr. Toyokawa for providing his code. 
\end{acknowledgments}

\bibliography{Folding-Pb}

\end{document}